\documentclass[aps,prl,floatfix,twocolumn,showpacs]{revtex4-1}
\usepackage[utf8x]{inputenc}
\usepackage{textcomp}
\usepackage{amsmath,amssymb,latexsym,graphicx}
\usepackage{graphics}       
\DeclareGraphicsExtensions{.png,.pdf}
\usepackage{mathtools}
\usepackage{overpic}
\usepackage{color}
\usepackage{subfigure}
\usepackage{natbib}
\usepackage[colorlinks=true,a4paper=true, pdfstartview=FitV,linkcolor=blue, citecolor=blue, urlcolor=blue]{hyperref}
\usepackage[all]{hypcap}
\usepackage{times}
\usepackage{multirow}
\usepackage{bm}
\usepackage{mathrsfs}
\usepackage{relsize}
\newcommand*{\diff}{\mathop{}\!\mathrm{d}}
\newcommand{\sgn}{\operatorname{sgn}}

\begin{document}
\title{Spin Density Wave Fluctuations and $p$-wave Pairing in Sr$_2$RuO$_4$}
\author{Jia-Wei Huo$^1$, T. M. Rice$^{2,1}$ and Fu-Chun Zhang$^{1,3}$}
\affiliation{
  $^1$Department of Physics and Centre of Theoretical and Computational Physics, The University of Hong Kong, Hong Kong SAR, China\\
$^2$Institut f\"{u}r Theoretische Physik, ETH Z\"{u}rich, CH-8093, Z\"{u}rich, Switzerland\\
$^3$Department of Physics, Zhejiang University, Hangzhou 310027, China
} 
\date{\today}
\begin{abstract}
  Recently a debate has arisen over which of the two distinct parts of the Fermi surface of Sr$_2$RuO$_4$, is the active part for the chiral $p$-wave superconductivity. Early theories proposed $p$-wave pairing on the two dimensional $\gamma$-band, while a recent proposal focuses on the one dimensional ($\alpha,\beta$) bands whose nesting pockets are the source of the strong incommensurate spin density wave (SDW) fluctuations. We apply a renormalization group theory to study quasi-one dimensional repulsive Hubbard chains and explain the form of SDW fluctuations, reconciling the absence of long range order with their nesting Fermi surface. The mutual exclusion of $p$-wave pairing and SDW fluctuations in repulsive Hubbard chains favors the assignment of the two dimensional $\gamma$-band as the source of $p$-wave pairing.
\end{abstract}

\maketitle

While there is a general consensus in favor of a chiral $p$-wave superconducting (SC) state in Sr$_2$RuO$_4$ \cite{Mackenzie2003,Bergemann2003,Maeno2012,Kallin2012} stabilized predominantly in either the 2-dimensional (2D) $\gamma$ band, or the pair of quasi-1D ($\alpha,\beta$) bands, a debate has arisen recently over which of the two is the active source of the superconductivity. The debate has been triggered by the failure to observe the persistent edge currents associated with the chirality, though this also has been controversial \cite{Maeno2012,Raghu2010a,Imai2012}. Microscopic derivations of a chiral $p$-wave pairing state in the 2D-scenario have been proposed based on a 2D Hubbard model by a $T$-matrix approach \cite{Hlubina1999}, third perturbation theory \cite{Nomura2000} and functional renormalization group (RG) calculations \cite{Honerkamp2003}. On the other hand, Raghu {\it et. al.}\cite{Raghu2010a,Raghu2012} have recently argued for the quasi-1D-scenario as more compatible with the missing edge currents and have provided a microscopic justification for it by using a RG theory, but only in the limit of weak interactions.

A closely related and competing phenomenon in Sr$_2$RuO$_4$ is the strong spin density wave (SDW) fluctuations at an incommensurate nesting wavevector spanning the Fermi surfaces of the 4/3 filled ($\alpha,\beta$) bands \cite{Braden2002}.  SDW fluctuations at this wavevector $\vec{\bm{Q}}=(2\pi/3,2\pi/3)$ \footnote{This wavevector is equivalent to $(4\pi/3,4\pi/3)$ in the electron notion.}  were recently reported at room temperature and at energies as high as 80 meV \cite{Iida2011}. The SDW peaks at $\vec{\bm{Q}}$ which combine nesting in both nearly 1D Fermi surfaces, grow as the temperature ($T$) is reduced and saturate at the crossover to 3D Fermi liquid behavior at $T_{3D}\approx$ 60 K \cite{Braden2002} when the resistivity starts to show a $T^2$ behavior \cite{Maeno1994}. Smaller peaks were observed at the wavevectors ($\pi,2\pi/3$) and ($2\pi/3,\pi$). To date these have been discussed within a random phase approximation (RPA) scheme \cite{Mazin1999,Kee2000,Morr2001,Eremin2002}. Because of the highly nesting character of the ($\alpha,\beta$) Fermi surface, it is necessary to choose a very weak interaction in RPA with a value typically an order of magnitude smaller than standard estimates \cite{Mravlje2011,Georges2012}.

In this Letter we show that treating the 1D character of the ($\alpha,\beta$) bands in a RG scheme, can explain the strong SDW fluctuation and reconcile the absence of the SDW long range order at $T>T_{3D}$ using a standard value for the interaction as observed in Sr$_2$RuO$_4$. Furthermore, our RG scheme shows mutual exclusion of $p$-wave pairing and SDW fluctuations in repulsive Hubbard chains, and a sharp suppression of the SDW fluctuations at low frequency in the $p$-wave SC state. Such a suppression has not been observed in previous neutron experiments and will be a challenge to explain within the quasi-1D-scenario.

We start from a 1D RG \cite{Solyom1979} treatment of the single chains in the ($\alpha,\beta$) bands. This includes the important cancellation between particle-hole and particle-particle graphs, which is absent in RPA. The properties of single chains in one-loop RG were derived in an early application of RG to condensed matter systems \cite{Solyom1979}. With repulsive interactions, the SDW and triplet superconductivity (TS) response functions have a power law form with divergences to infinity or zero, as $T\to 0$. The phases with enhanced SDW and suppressed TS and vice versa, are separated by a Quantum Critical Point where the exponent $\theta$ changes sign. In the one-loop approximation, $\theta=g_2-g_1/2$, where $g_1$ ($g_2$) is the dimensionless scattering coefficient for backward (forward) processes. The response functions in 1D RG behave differently from that in RPA, which gives a finite scale divergence in the SDW response. This RPA scale corresponds to a large temperature (up to $10^3$ K) in Sr$_2$RuO$_4$ if we use typical values of the intrachain hopping $t=0.3$ eV \cite{Bergemann2003} and Hubbard repulsion $U=2.2$ eV \cite{Mravlje2011,Georges2012}.

To begin with, we consider the hybridization and interaction between the two orbitals, $d_{xz}$ and $d_{yz}$, which give rise to perpendicular chains. To obtain estimates of the mean field transition temperature to SDW order in Sr$_2$RuO$_4$ we use the results of numerical calculations \cite{Hirsch1983, Imada1989} to extend the one-loop RG calculations to stronger interactions. Next, we examine the interactions between parallel chains. This leads to a theory, which gives a strong SDW fluctuation without long range order at low temperatures.

The effect of the hybridization and spin-orbit coupling among the two orbitals can be described by the following Hamiltonian \cite{Raghu2010a}
\begin{multline}
  \delta H=\sum_{\vec{k},\sigma}(-2t''\sin k_x\sin k_y c^\dag_{\vec{k},xz,\sigma}c_{\vec{k},yz,\sigma}+\text{H.c.})\\
  +\eta\sum_{m,n}\sum_{\sigma\sigma'}\sum_{\vec{k}}c^\dag_{\vec{k},m,\sigma}c_{\vec{k},n,\sigma'}\vec{\ell}_{mn}\cdot\vec{\sigma}_{\sigma\sigma'}.
\end{multline}
Here $c^\dag_{\vec{k},m,\sigma}$ is the electronic creation operator in the orbital $m$ with momentum $\vec{k}$ and spin $\sigma$, and the angular momentum operators and spin operators are represented in terms of the totally anti-symmetric tensor $\ell^a_{mn} = i \epsilon_{a mn}$ and Pauli matrices $\vec{\sigma}$, respectively. For the system we are interested in, the strengths of the mixing and spin-orbit coupling are $t''\approx 0.1t$ and $\eta \approx 0.1t $, respectively \cite{Ng2000,Kontani2008}.

When the above perturbation $\delta H$ is taken into account, the quasi-particle spectrum opens a gap $2\lambda$ near $(\pm k_F,\pm k_F)$ with $\lambda=\sqrt{(3t''/2)^2+\eta^2}$. Therefore, the dispersion for the $d_{xz}$ orbital is modified to
\begin{equation}
\epsilon_{\vec{k}}\approx v_F(|k_x|-k_F)+\sgn(|k_x|-k_F)L(|k_y|)\lambda,  
\end{equation}
with $L(x)$ the Lorentzian function centered at $k_F$. The Green's function describing quasi-particle excitations in the $d_{xz}$ orbital is $G_{xz}(k_x,k_y,i\omega_n)=1/(i\omega_n-\epsilon_{\vec{k}})$, and a similar counterpart in the $d_{yz}$ orbital. Then the bare SDW response function reads
\begin{equation}
  \chi^{00}_{xz}(\vec{\bm{q}},i\Omega)=2\frac{T}{N}\sum_{\vec{k},i\omega_n}G_{xz}(\vec{k}+\vec{\bm{q}},i\omega_n+i\Omega)G_{xz}(\vec{k},i\omega_n),
\end{equation}
with $N$ the total number of sites, and we work in units where the lattice spacing is unity. For $\vec{\bm{q}}=(2k_F,q_y)$, we have, after a analytic continuation to real frequency with $\omega=0$,
\begin{multline}
  \chi^{00}_{xz}(\vec{\bm{q}},T)=\frac{1}{\pi v_F}\left[\ln\frac{4T}{E_0+2\lambda}+\right.\\
  \left.\int_0^\infty\ln\left(x+\frac{\lambda}{2T}\right)\text{sech}^2x\diff x\right],
\end{multline}
with $E_0=4t$ the band width and $v_F$ the Fermi velocity, where we have set $\hbar=1$. The bare SDW can be approximated by the following interpolation scheme, in the limit of small $\lambda$,
\begin{equation}
  \chi^{00}_{xz}(\vec{\bm{q}},T)=\chi^{00}(\vec{\bm{q}},T)\approx\frac{1}{\pi v_F}\ln\frac{T+2\lambda}{E_0}.
\end{equation}

A standard RG calculation, by including the particle-particle and particle-hole graphs, gives the dressed susceptibility in terms of intraorbital interactions:

\begin{equation}\label{eqn:bare_chi}
  \chi^{0}_{xz}(\vec{\bm{q}},T)=\chi^0(\vec{\bm{q}},T)=\frac{1}{\pi v_F\theta}\left(\frac{E_0}{T+2\lambda}\right)^\theta,
\end{equation}
with $\theta=U/2\pi v_F$ in the Hubbard model. To deal with the case of strong interactions, we compared our results with the $T$-dependence of spin susceptibility from the Monte Carlo method\cite{Hirsch1983, Imada1989}, and found that $\theta$ is screened to be $0.6\theta$, which we denote as $\theta^*$ from now on. From the expression above, one can see that due to the hopping between the two orbitals, a finite low energy cut-off $\lambda$ appears, killing the divergence as $T\rightarrow 0$.

Next we introduce the inter-orbital interactions between the two $4d$ orbitals, with the generic form of the on-site interaction\cite{Mravlje2011}
\begin{multline}\label{eqn:interaction}
  H_I=\sum_mUn_{m\uparrow}n_{m\downarrow}+\sum_{m<n,\sigma}[U'n_{m\sigma}n_{n\bar{\sigma}}+(U'-J_H)n_{m\sigma}n_{n\sigma}\\
  -J_Hc^\dag_{m\sigma}c_{m\bar{\sigma}}c^\dag_{n\bar{\sigma}}c_{n\sigma}]-J_H\sum_{m<n}[c^\dag_{m\uparrow}c^\dag_{m\downarrow}c_{n\uparrow}c_{n\downarrow}+\text{H.c.}],
\end{multline}
with $U'$ the interorbital Coulomb repulsion and $J_H$ the Hund’s rule coupling, in addition to the intraorbital repulsion $U$. Here $c^\dag_{m\sigma}$ is the creation operator for electrons in the orbital $m$ with spin $\sigma$ at the same site, and $n_{m\sigma}$ is the corresponding number operator. Here we note that the intraorbital interaction $U$ has been taken into account in the exponent of Eq.~\ref{eqn:bare_chi}.

To incorporate the above interactions, we define the joint response function for SDW, by including the orbital indices $m$,
\begin{equation}\label{eqn:joint}
  \chi_H(\vec{\bm{q}},i\Omega)=-\int_0^\beta e^{i\Omega\tau}\langle T_\tau\mathcal{O}(\vec{\bm{q}},\tau)\mathcal{O}^\dag(\vec{\bm{q}},0)\rangle\diff\tau,
\end{equation}
where
\begin{multline}
  \mathcal{O}(\vec{\bm{q}},\tau)=\frac{1}{\sqrt{N}}\sum_{\vec{k},m}[c^\dag_{\vec{k},m\uparrow}(\tau)c_{\vec{k}+\vec{q},m\downarrow}(\tau)-\\c^\dag_{\vec{k},m\downarrow}(\tau)c_{\vec{k}+\vec{q},m\uparrow}(\tau)],  
\end{multline}
with $\vec{\bm{q}}=(2k_F,q_y)$ and $(q_x,2k_F)$ for $d_{xz}$ and $d_{yz}$ orbitals, respectively.

\begin{figure}[ht]\centering
  \vspace{5pt}
  \begin{tabular}{c}
    \resizebox{0.8\linewidth}{!}{
      \begin{overpic}{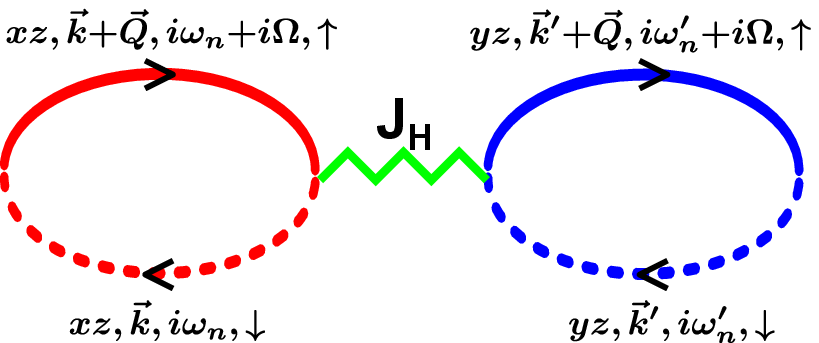}
        \put(-13,35){\resizebox{15pt}{!}{(a)}}
      \end{overpic}} \\
    \resizebox{0.85\linewidth}{!}{
      \begin{overpic}{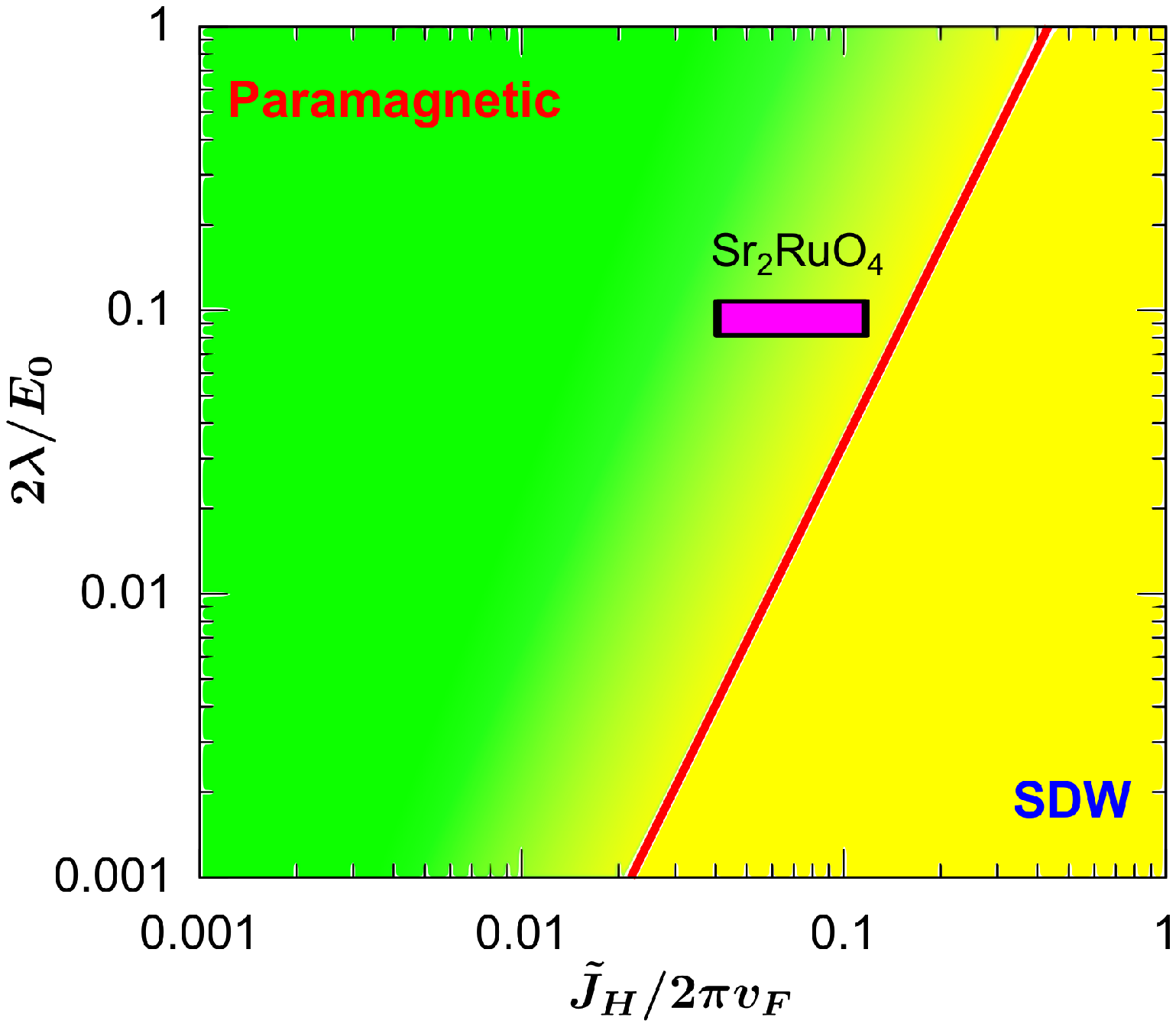}
        \put(-7.7,75){\resizebox{28pt}{!}{(b)}}
      \end{overpic}}
  \end{tabular}
\caption{\label{coupling}(Color online) (a) The lowest order Feynman diagram for the spin-spin correlation function connecting propagators from different orbitals via the Hund's rule coupling. The red and blue lines stand for electrons in the $xz$ and $yz$ orbitals, respectively. The solid and dashed lines correspond to electrons belonging to the branches containing $+k_F$ and $-k_F$ in the one-dimensional model, respectively. Due to the conservation of momentum at the vertex, $\vec{\bm{Q}}$ is locked to be $(2k_F,2k_F)$. (b) SDW vs. paramagnetic phase diagram in the parameter space of $\lambda$ and $\tilde{J}_H$ at $T=0$ from Eq.~\ref{eqn:tc}. The estimated parameter region for Sr$_2$RuO$_4$ is indicated in the paramagnetic phase. In this region, $\tilde{J}_H$ ranges from 0.13 eV to 0.4 eV, the renormalized and bare values of the Hund's rule coupling.}
\end{figure}

To first order in the inter-orbital interactions, the only non-vanishing term is
\begin{multline}
  J_H\frac{T^2}{N^2}\sum_{\vec{k},\vec{k}',i\omega_n,i\omega_n'}G_{xz}(\vec{k}+\vec{\bm{Q}},i\omega_n+i\Omega)G_{xz}(\vec{k},i\omega_n)\times\\
  G_{yz}(\vec{k}'+\vec{\bm{Q}},i\omega_n'+i\Omega)G_{yz}(\vec{k}',i\omega_n'),
\end{multline}
with a corresponding diagram shown in Fig.~\ref{coupling} (a). Note that the wave vector for the response function $\vec{\bm{Q}}=(2k_F,2k_F)$ is the same for both $d_{xz}$ and $d_{yz}$ orbitals due to the conservation of momentum in the scattering process in Fig.~\ref{coupling}. Another important consequence is that only the Hund's rule coupling contributes to the SDW response function, while other onsite interaction terms in Eq.~\ref{eqn:interaction} are not involved. This result originates from the spin configuration in Fig.~\ref{coupling}. In this sense, the Hund's rule coupling assists the spin-flip processes between different orbitals. An intuitive physical picture is that the spin-flip processes are coherent even in different orbitals, due the ferromagnetic Hund's rule coupling between the two orbitals. Dynamical mean-field theory found that the Hund's rule coupling is important in Sr$_2$RuO$_4$ \cite{Mravlje2011}. In our calculations below we use a Gutzwiller renormalized values of $\tilde{J}_H\approx 0.13$ eV to take into account the strong onsite repulsion between holes \footnote{The effect of strong onsite repulsion suppresses the probability of having two electrons from distinct orbitals at the same site to be 1/3. Thus, $\tilde{J}_H$ is reduced by a Gutzwiller factor of 1/3, compared with the bare value $J_H=0.4$ eV in Ref.~\cite{Mravlje2011}.}.

The full dressed joint SDW response function in Eq.~\ref{eqn:joint} is obtained by first including the intra-orbital interaction $U$ in a RG scheme, which means that the bare bubbles in Fig.~\ref{coupling} (a) are replaced with the dressed ones in Eq.~\ref{eqn:bare_chi}. Due to the absence of the low energy divergence in Eq.~\ref{eqn:bare_chi}, the Hund's rule coupling $\tilde{J}_H$ can be treated in a RPA-like method, leading to
\begin{equation}\label{eqn:chi}
  \chi_H(\vec{\bm{Q}},T)=\frac{2\chi^0(\vec{\bm{Q}},T)}{1-\frac{\tilde{J}_H}{2}\chi^0(\vec{\bm{Q}},T)},
\end{equation}
The divergence of $\chi_H(\vec{\bm{Q}},T)$ in Eq.~\ref{eqn:chi} gives estimate of the mean field transition temperature to long range SDW order,
\begin{equation}\label{eqn:tc}
  T^{\text{SDW}}_c=E_0\left(\frac{\tilde{J}_H}{2\pi v_{F}\theta^*}\right)^{\frac{1}{\theta^*}}-2\lambda.
\end{equation}
The first term gives an upper limit on $T^{\text{SDW}}_c$ due to the Hund's rule coupling, which is about 50 K, similar to $T_{3D}$. The presence of the second term, of order $10^3$ K, guarantees that the ground state is paramagnetic. To illustrate the underlying physics, we construct the phase diagram in the $\tilde{J}_H-\lambda$ parameter space as shown in Fig.~\ref{coupling} (b). The parameter set for Sr$_2$RuO$_4$ is in the paramagnetic region, but near the phase boundary with SDW fluctuations. Therefore, the strongly enhanced response function by the Hund's rule coupling at $\vec{\bm{Q}}$ naturally explains the strong enhancement of the SDW signal near $\vec{\bm{Q}}$ in the experiments \cite{Braden2002}.

Next, we briefly study how the interchain tunnelling between parallel chains affects the SDW in the quasi-one dimensional ($\alpha,\beta$) bands of Sr$_2$RuO$_4$. The interchain hopping $t_\bot$ alone would give rise to a singular SDW response function at the wavevector ($2k_F$,$\pi$) since the tight-binding approximation preserves the perfect nesting for quasi-one dimensional systems \cite{Horovitz1976, Solyom1979}. However, in the case of Sr$_2$RuO$_4$, the Fermi surfaces the ($\alpha,\beta$) bands are distorted due to hybridization and spin-orbit coupling between orbitals, as discussed previously. Therefore, the nesting property at ($2k_F$,$\pi$) is lost, and a strong enhancement for the SDW fluctuation is not expected.

Another mechanism to affect the SDW response function at ($2k_F$,$\pi$) is the superexchange interaction $J_{\text{ex}}=4t^2_\bot/U$ between two neighboring parallel chains. But a rough estimation yields $J_{\text{ex}}\ll J_H$, since $t_\bot$ is only about 0.026 eV \cite{Bergemann2003}. Combining the above two effects for the parallel chains, the spin fluctuation response at ($2k_F,\pi$) should be much weaker than that at ($2k_F,2k_F$), as observed in the experiment \cite{Iida2011}.


Finally, we consider the effects due to the SC pairing in 1D bands on the magnetic response. Earlier measurements did not show a change in the magnetic response at $\vec{\bm{Q}}$ upon cooling through the SC transition at $T^s_c$($=1.5$ K) \cite{Braden2002}.

To make our analysis more transparent, we restrict the discussion to one dimension, and thus consider the following Hamiltonian $H=H_0+H_{\text{int}}$, where $H_{\text{int}}=U\sum_{i}n_{i\uparrow}n_{i\downarrow}$ is the Hubbard onsite interaction term, which can be reduced to the standard form describing different scattering processes with $g_1\!=\!g_2\!=\!U/\pi v_F$ \cite{Giamarchi2003,Solyom1979}. To incorporate the SC pairing, we consider $H_0$ as
\begin{multline}
  H_0=\sum_{k,\sigma}v_F(|k|-k_F)c^\dag_{k\sigma}c_{k\sigma}\\+\sum_{k}[\Delta(k)c^\dag_{k\uparrow}c^\dag_{-k\downarrow}+\text{H.c.}],
\end{multline}
which models a one-dimensional electrons with $p$-wave SC pairing $\Delta(k)$ and can be solved exactly in the mean-field approximation. The assumption that spin-orbit coupling locks the $\vec{d}$ vector along the crystal $c$ axis has been made, consistent with the polarized-neutron scattering experiment in Sr$_2$RuO$_4$ \cite{Duffy2000}.

A powerful tool to study the properties of $H_0$ is the application of the normal and anomalous Green's functions $G_{\sigma\sigma'}(k,i\omega_n)$ and $F_{\sigma\sigma'}(k,i\omega_n)$ \cite{Abrikosov1975}, which read
\begin{equation}
  G_{\sigma\sigma'}(k,i\omega_n)=-\delta_{\sigma\sigma'}\frac{i\omega_n+\xi_k}{\omega^2_n+\xi_k^2+\Delta_0^2}
\end{equation}
and
\begin{equation}
  F_{\sigma\sigma'}(k,i\omega_n)=\frac{\bm{\Delta}_{\sigma\sigma'}(k)}{\omega^2_n+\xi_k^2+\Delta_0^2}.
\end{equation}
Here $\xi_k=v_F(|k|-k_F)$ and $\bm{\Delta}_{\sigma\sigma'}(k)=\Delta(k)\sigma_zi\sigma_y$. Near the Fermi surface, we have, due to the odd parity, $\Delta(k)=\sgn(k)\Delta_0$, with $\Delta_0$ the SC gap near the Fermi surface.

Interestingly, after some algebra, we find that the particle-particle and particle-hole diagrams just differ by a minus sign, similar to the case without SC pairing. For example, the particle-hole bubble diagram can be expressed as
\begin{multline}
  \frac{T}{N}\sum_{k,i\omega_n}G_{\uparrow\uparrow}(k,i\omega_n)G_{\downarrow\downarrow}(k+2k_F,i\omega_n+i\Omega)-\\
  F_{\uparrow\downarrow}(k,i\omega_n)F_{\downarrow\uparrow}^\dag(k+2k_F,i\omega_n+i\Omega).
\end{multline}
To the leading order in the logarithmic accuracy, this expression is reduced to
\begin{equation}
  \frac{1}{2\pi v_F}\left[\ln\frac{\sqrt{|\omega^2-4\Delta_0^2|}}{E_0}-i\frac{\pi}{2}\Theta(\omega-2\Delta_0)\right],
\end{equation}
with $\Theta(x)$ the Heaviside function, and we have performed an analytic continuation to real frequency $\omega$ at zero temperature. The structure of this expression is also similar to its counterpart in the normal state. And due to this analogy, the RG flow equations for the interaction constants $g_1$ and $g_2$ should be the same as the non-SC case \cite{Giamarchi2003,Solyom1979}. Therefore, in the case of $\theta^*>0$, the SDW phase is expected as usual.

\begin{figure}[ht]\centering
  \begin{tabular}{c}
    \resizebox{\linewidth}{!}{
      \begin{overpic}{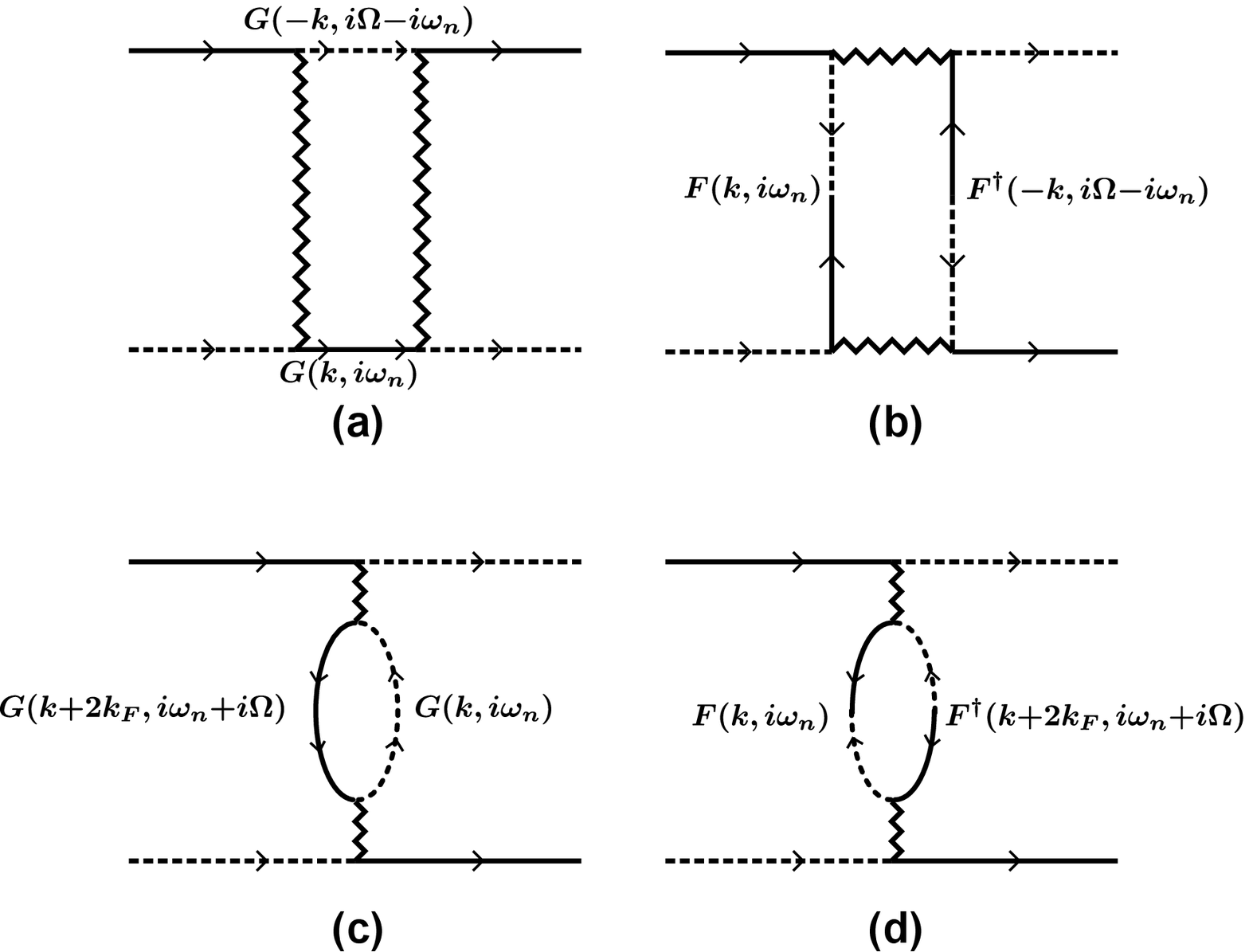}
      \end{overpic}} \\
    \resizebox{0.7\linewidth}{!}{
      \begin{overpic}{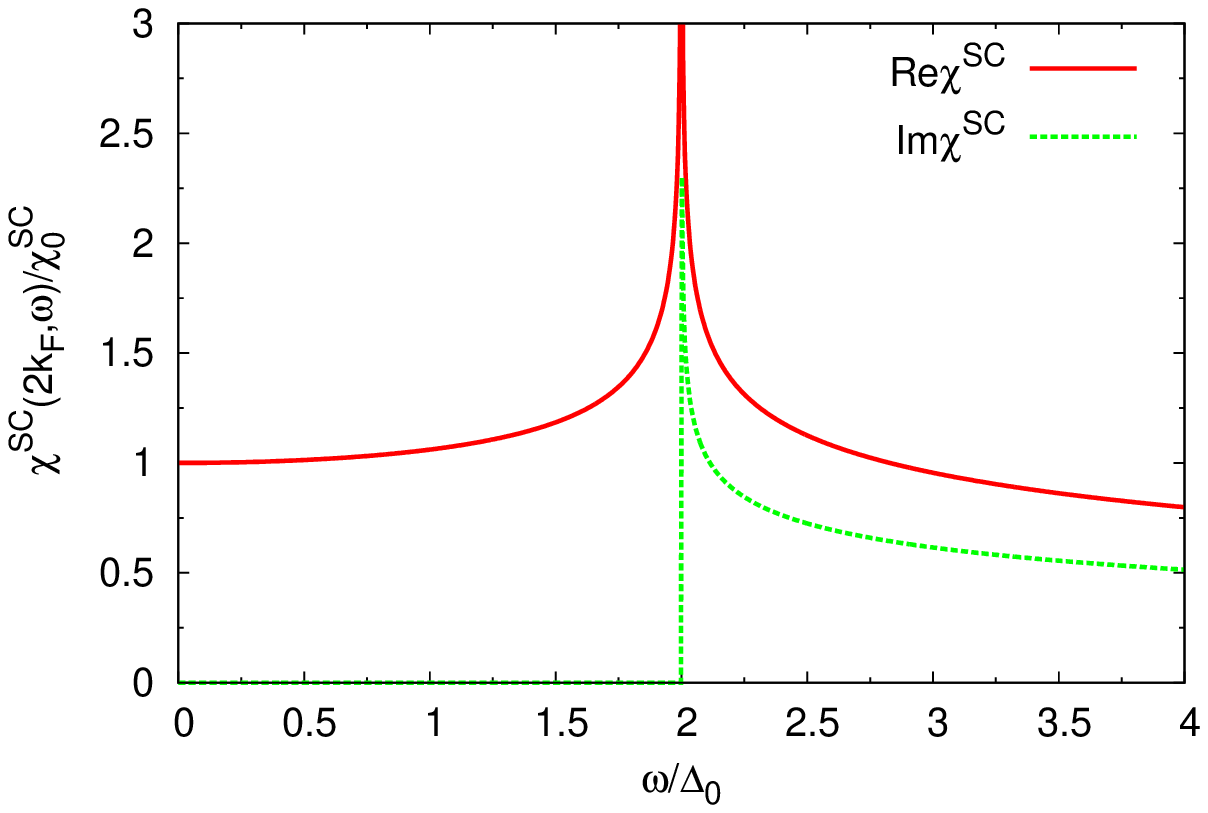}
        \put(80,15){\resizebox{22pt}{!}{(e)}}
      \end{overpic}}
  \end{tabular}
\caption{\label{sc}(Color online) (Upper panels) Structure of the vertex diagrams. (a) and (b) are for the particle-particle channel, whereas (c) and (d) are for the particle-hole channel. The solid and dashed lines correspond to electrons belonging to the branches containing $+k_F$ and $-k_F$ in the one-dimensional model, respectively. The wavy lines stand for bare onsite interactions. (Lower panel) Response function as a function of $\omega$ at $T=0$ is shown in (e). Here $\chi^{\text{SC}}(2k_F,\omega)$ is scaled by $\chi_0^{\text{SC}}\equiv\text{Re}\chi^{\text{SC}}(2k_F,\omega=0)$, and $\theta^*$ is about 0.41 in the system of interest.}
\end{figure}

A standard RG analysis yields the final results for the response function in the SC state as follows:
\begin{eqnarray}
  \text{Re}\chi^{\text{SC}}(2k_F,\omega)&=&\frac{\left(\frac{E_0}{\sqrt{|\omega^2-4\Delta^2_0|}}\right)^{\theta^*}}{\pi v_F\theta^*}\nonumber\\
  \text{Im}\chi^{\text{SC}}(2k_F,\omega)&=&\frac{\Theta(\omega-2\Delta_0)}{2v_F}\left(\frac{E_0}{\sqrt{|\omega^2-4\Delta^2_0|}}\right)^{\theta^*}.
\end{eqnarray}
As is shown in Fig.~\ref{sc} (b), if one looks at the low-energy properties $\omega\to 2\Delta_0$, the response function diverge as $\chi\sim|\omega-2\Delta_0|^{-\theta^*/2}$. This result indicates that the transition to superconductivity in the 1D bands will open a gap in the low energy spectra at wavevector $\vec{\bm{Q}}$ \cite{Kee2000,Morr2001}. While early neutron scattering experiments by Braden and coworkers \cite{Braden2002} did not show a change in low energy spectra at $\vec{\bm{Q}}$, a more complete investigation would be worthwhile to definitively decide if an SC gap opens up in the 1D ($\alpha,\beta$) bands at the onset of superconductivity at $T^s_c=1.5$ K.

In summary, we have applied a RG scheme starting from the 1D analysis for single chains, to explain the strong SDW fluctuations and the absence of SDW order at temperature above the crossover to 3D Fermi liquid behavior with the strong onsite Hubbard repulsion estimated for Sr$_2$RuO$_4$. The mutual exclusion in 1D RG theory of enhancement in the SDW and simultaneously in the $p$-wave pairing channel is in favor of the 2D $\gamma$-band as the source of the superconductivity.

We thank M. Sigrist for useful discussions. JWH and FCZ wish to thank S. Raghu for interesting and helpful discussions. This work is partly supported by Hong Kong's RGC grant GRF HKU707211. T.M.R. acknowledges support from the Swiss Nationfonds.

\bibliography{mylib}
\bibliographystyle{apsrev4-1}
\end{document}